\documentclass[,]{aa}

\usepackage{txfonts}

\usepackage{epsfig,graphicx,amssymb,natbib}

\bibpunct{(}{)}{;}{a}{}{,} 

\title{Gamma-ray bursts and X-ray melting of material to form chondrules and planets.}

\titlerunning{GRBs and X-ray melting of material to form chondrules and planets.  }

\author{    P.\,Duggan\inst{1}      \and
            B.\,McBreen\inst{1}     \and
            A.\,J.\,Carr\inst{2}    \and
            E.\,Winston\inst{1}     \and
            G.\,Vaughan\inst{3}     \and
            L.\,Hanlon\inst{1}      \and
            S.\,McBreen\inst{1}     \and
            L.\,Metcalfe\inst{4}    \and
            \AA.\,Kvick\inst{3}     \and
            A.\,E.\,Terry\inst{3}
}

\institute{
    Department of Experimental Physics, University College Dublin, Dublin 4, Ireland.    \and
    Mechanical Engineering Department, University College Dublin, Dublin 4, Ireland.     \and
    European Synchrotron Radiation Facility, B. P. 220, F - 38043 Grenoble Cedex, France. \and
    XMM-Newton science Operations Center, European Space Agency,
Villafranca del Castillo,  28080 Madrid, Spain. }

\offprints{brian.mcbreen@ucd.ie}

\date{Received  19 June 2003 / Accepted 8 August 2003}

\abstract{Chondrules are millimeter sized objects of spherical to
irregular shape that constitute the major component of chondritic
meteorites that originate in the region between Mars and Jupiter
and which fall to Earth. They appear to have solidified rapidly
from molten or partially molten drops. The heat source that melted
the chondrules remains uncertain. The intense radiation from a
gamma-ray burst (GRB) is capable of melting material at distances
up to 300 light years. These conditions were created in the
laboratory for the first time when millimeter sized pellets were
placed in a vacuum chamber in the white synchrotron beam at the
European Synchrotron Radiation Facility (ESRF).  The pellets were
rapidly heated in the X-ray and gamma-ray furnace to above
1400\,\degr C melted and cooled. This process heats from the
inside unlike normal furnaces. The melted spherical samples were
examined with a range of techniques and found to have
microstructural properties similar to the chondrules that come
from meteorites. This experiment demonstrates that GRBs can melt
precursor material to form chondrules that may subsequently
influence the formation of planets. This work extends the field of
laboratory astrophysics to include high power synchrotron sources.
     }

\newcommand{\up}[1]{\raisebox{1.5ex}[0cm][0cm]{#1}}

\begin{document}

\maketitle

\section{Introduction}
The production of chondrules is an important stage in processes
leading to the formation of planets \citep{cuzzi2001}. Many
proposals \citep{boss1996} have been made for the transient heat
source that melted the precursor material to form chondrules
including nebular lightning \citep{desch2000} and activities
associated with the young Sun \citep{shu2001,feigelson2002}.
Almost all  proposed heat sources
\citep{rubin2000,boss1996,jones2000} are local to the solar
nebula, one exception being the proposal
\citep{mcbreen1999,duggan2001} that the chondrules were flash
heated to melting point by a nearby GRB when the iron all through
the precursor material efficiently absorbed X-rays and low energy
$\gamma$-rays. The distance to the source could be up to 300 light
years ($\sim$\,100\,pc) for a GRB with an isotropic output of
10$^{46}$\,J \citep{piran1999,mes2002}. This distance limit was
obtained using the minimum value of 2\,x\,10$^6$\,J\,kg$^{-1}$
required to heat and melt the precursor grains \citep{wasson1993}
and is equivalent to an enormous energy deposition of
10$^8$\,J\,m$^{-2}$. The complex time profiles of short and long
GRBs are well described by lognormal distributions
\citep{mcbreen2001,quilligan2002}.

\begin{figure}[t]\centering
\includegraphics[width=.82\columnwidth]{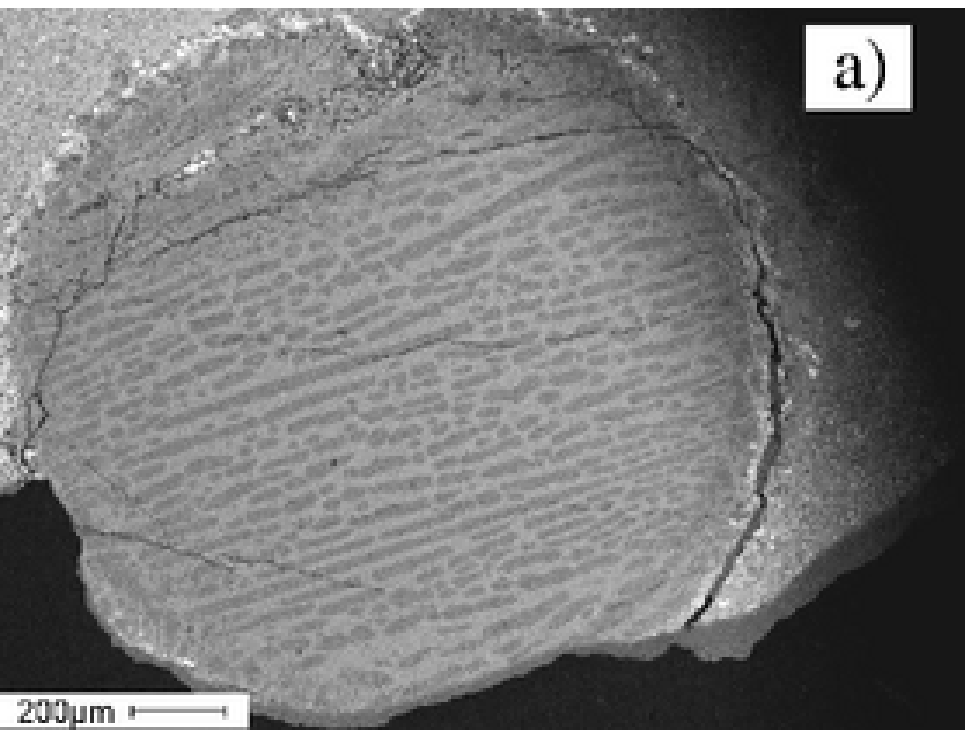}\\[3pt]
\includegraphics[width=.82\columnwidth]{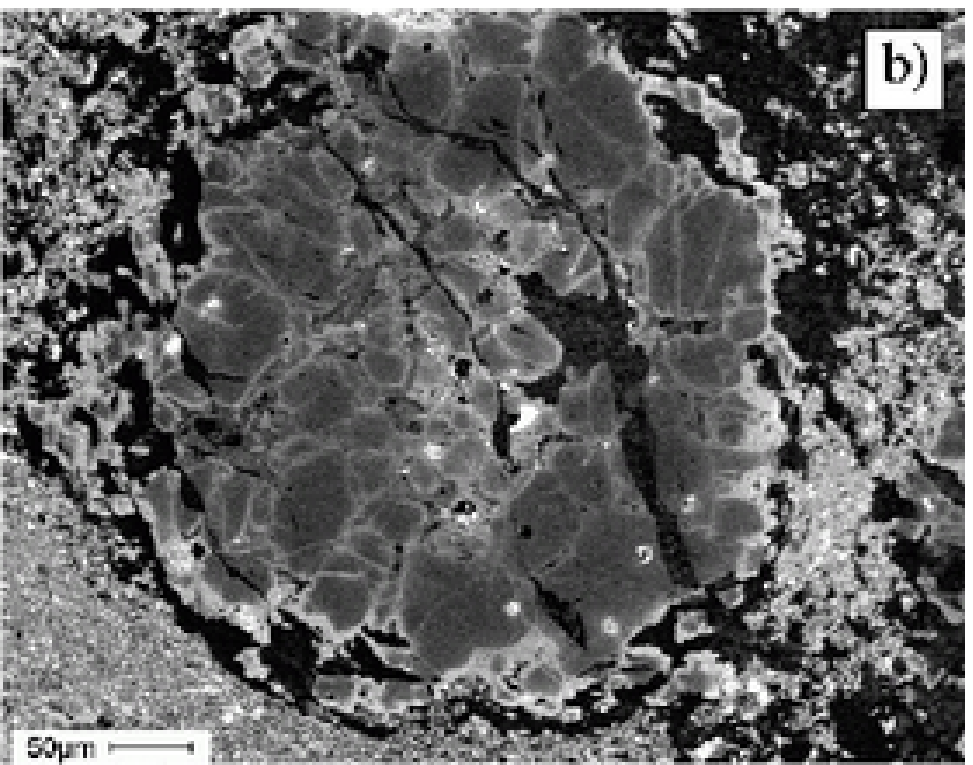}
    \caption{Backscattered electron microscope images of two
chondrules from the Allende meteorite. Phases with higher atomic
number are brighter in color. The dark grey grains are elongated
olivine crystals in (a) and porphyritic crystals in (b) where the
brighter regions are the interstitial glassy material. The
diameters of the chondrules are about 1\,mm and 0.6\,mm and are
surrounded by matrix material in the meteorite.}
    \label{fig:met}
\end{figure}

The timing, duration and conditions that led to the formation of
chondrules in the solar nebula are poorly understood. The
properties and thermal histories of the chondrules have been
inferred from an extensive series of experiments
\citep{connolly1998,lofgren1990,connolly1994,cohen2000}. Their
mineralogy is dominated by olivine ((Mg,Fe)$_2$SiO$_4$) and
pyroxene ((Mg,Fe)SiO$_3$) both being solid solutions with a wide
range in composition. This diversity is consistent with the
melting of heterogeneous solids or dust balls. Crystal structures
and morphologies are used to limit the temperature range and rate
of cooling during chondrule formation. Backscattered electron
microscope images of two typical chondrules that we obtained from
the Allende meteorite with barred and porphyritic textures are
shown in Fig.\,\ref{fig:met}.

\begin{table}[t]\centering
\caption{Elemental oxide compositions of three types of precursor
material in weight percentages and the calculated liquidus
temperatures. The compositions are the same as listed
\citep{yu1998} except that the small amounts of K$_2$O
($<$\,0.11\,\%) were not included and magnetite (Fe$_3$O$_4$) was
used in place of FeO.} \label{tab:comp}\small
\begin{tabular}{lccc} \hline\hline
            & \underline{Type\,IA}& \underline{Type\,IAB} &
\underline{Type\,II} \\
SiO$_2$         &   45.4   & 47.2  &  49.2  \\
TiO$_2$         &   0.1    & 0.1   &  0.1   \\
Al$_2$O$_3$     &   4.9    & 9.7   &  4.1   \\
Fe$_3$O$_4$     &   8.3    & 6.6   &  21.3  \\
MnO             &   0.1    & 0.1   &  0.1   \\
MgO             &   37.4   & 30.7  &  22.7  \\
CaO             &   2.5    & 3.5   &  0.2   \\
Na$_2$O         &   1.3    & 2.1   &  2.3
\\[2pt]
Liquidus   \\
\,\,Temperature     & \up{1,692\,\degr C}     &\up{1,577\,\degr C}
&
\up{1,509\,\degr C}\\
\hline\hline
\end{tabular}\normalsize
\end{table}

\section{Experimental Method}
It is now possible to create the astrophysical conditions near a
GRB source in the laboratory due to the development of powerful
synchrotrons. The ESRF has a 6\,GeV, third-generation synchrotron
capable of generating the required power. A wiggler device was
inserted and used to create X-rays in the range 3 -- 200\,keV. The
24-pole wiggler has a characteristic energy of 29 keV at a minimum
wiggler gap of 20.3\,mm. Time was awarded on the ID 11 white beam
to test the prediction that large fluxes of X-rays and
$\gamma$-rays could melt millimetre sized dust grains and hence
extend the use of high power synchrotron sources to laboratory
astrophysics \citep{remington1999}.

The composition of chondrules varies widely and a classification
system based on the iron content is often used
\citep{mcsween1977,yu1998}. Type IAB chondrules have low iron
content while type IA and type II have increasing amounts of iron.
The pellets were made from a mixture of elemental oxides with
weight percentages as given in Table\,\ref{tab:comp} for the three
precursors types. The major effect of including more iron is to
reduce the magnesium content and the liquidus temperature of the
material. The oxides, without the volatile Na$_2$O, were mixed and
heated to 400\,\degr C in an alumina crucible for 13 hours. The
powders after heating had a mass loss of about 5\,\%, due to
moisture loss and reduction of the elemental oxides. After
cooling, Na$_2$O was added to the mixture. The powder was pressed
into cylindrical pellets of diameter 3\,mm and height of 3\,mm.

\begin{figure}[t]\centering
\includegraphics[width=0.8\columnwidth]{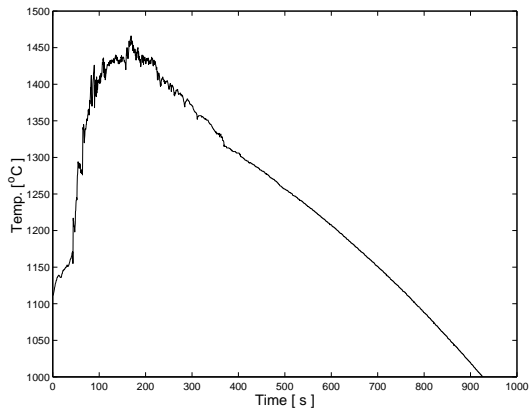}
    \caption{The temperature profile of a typical sample in the
synchrotron experiment. The reason for the high initial
temperature is that 1000\,\degr C is the minimum value read by the
pyrometer.  The temperature rose rapidly to above 1000\,\degr C
when the beam shutter was opened.}
    \label{fig:temp_profile}
\vspace*{-0.25cm}\end{figure}

Each pellet was placed in a graphite crucible inside an evacuated
container and inserted into the path of the white X-ray beam. The
size of the beam was 2\,mm x 1.5\,mm. The synchrotron beam entered
the vacuum chamber through a Kapton window of thickness 0.05\,mm.
The pressure in the container was between 10$^{-2}$ --
10$^{-3}$\,mbar, which is typical of planetary forming systems. In
a few cases the residual air in the vacuum chamber was replaced
with hydrogen. During the heating cycle the temperature of the
pellet was measured using a Raytex MR1SCSF pyrometer with a range
from 1000\,\degr C to 3000\,\degr C. The pyrometer was located
outside the lead shielding and viewed the sample via a mirror
through a glass window on the top of the vacuum container and at
right angles to the X-ray beam. This window had to be replaced on
several occasions during the experiment because of darkening
caused by radiation damage. The sample was also monitored with a
camera that viewed it through the pyrometer optics. The pellets
were rapidly heated in the X-ray and $\gamma$-ray furnace to
temperatures above 1400\,\degr C (Fig.\,\ref{fig:temp_profile}).
During the heating and melting process the pellets bubbled, moved
about and sometimes ejected small drops of iron rich material. The
melted samples were kept at the maximum temperature for a duration
of 10\,s to 300\,s and cooled when the power in the beam was
reduced by widening the magnets of the wiggler. The beam was
removed when the temperature dropped below 1000\,\degr C. The
samples cooled rapidly to yield 2\,--\,3\,mm diameter black
spherules.

A Huber model 642 Guinier X-ray powder diffractometer with
monochromatic Cu\,K$\alpha_1$ radiation was used for powder
diffraction. Aluminium foil was placed between the sample and
detector to reduce the fluorescent background from iron. The
crystal phases were identified using the JCPDS Powder Diffraction
File. Crystal structures were refined using Rietveld analysis in
the range 20$\degr\!<\!2\theta\!<\!100\degr$ using Rietica
\citep{hunter1998,hill1987}. For angular calibration and
quantitative phase analysis a known mass of high-purity silicon
(9\,\% to 16\,\% by mass depending on the sample) was added as an
internal standard to the powdered sample. Refined structural
variables included lattice parameters, atomic coordinates, metal
site occupancies and isotropic temperature factor; nonstructural
variables included scale factors, background correction and peak
shape. Absorption effects were treated as though they were part of
the overall isotropic temperature factor \citep{scott1981}.

\section{Results}

\begin{figure}[t]
\centering
\includegraphics[width=.82\columnwidth]{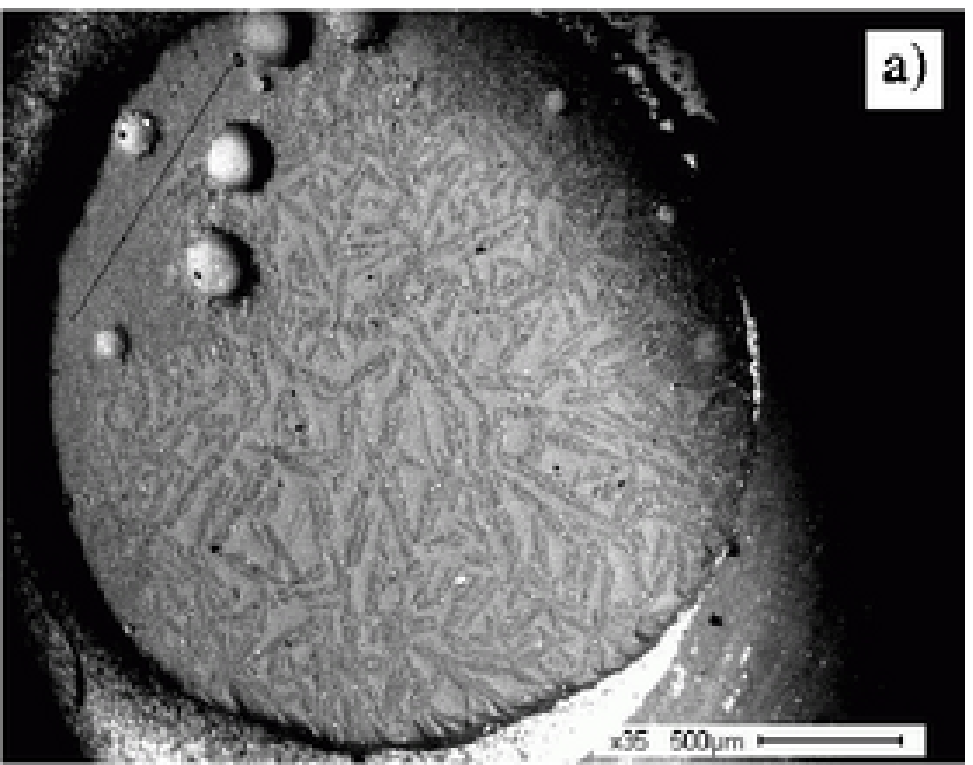}\\[1pt]
\includegraphics[width=.82\columnwidth]{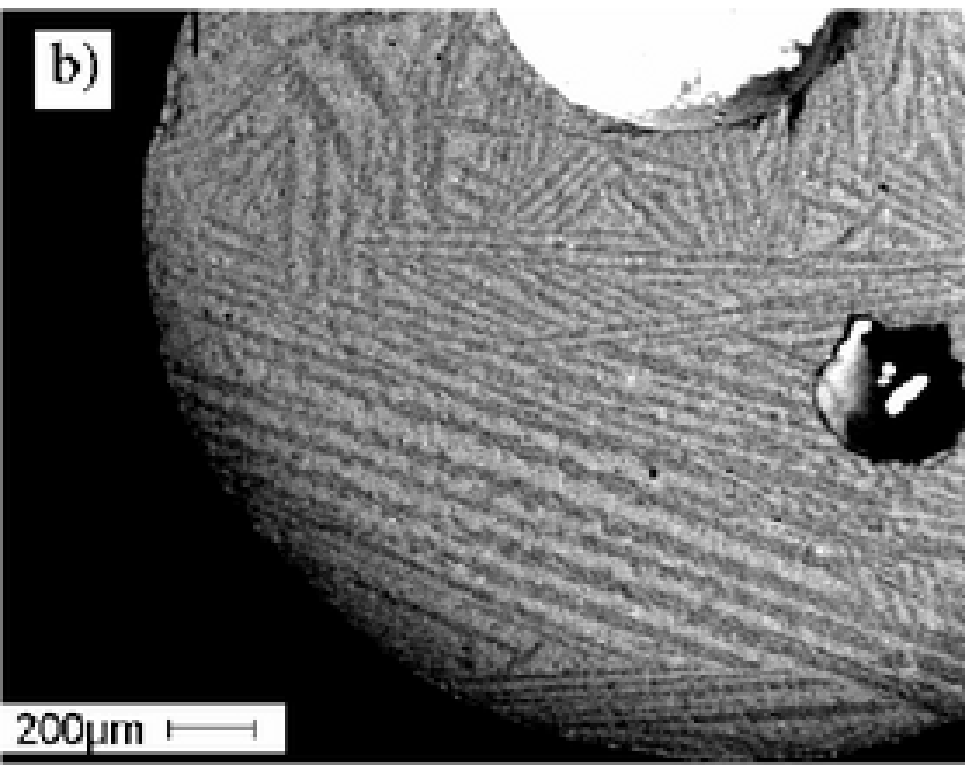}\\[1pt]
\includegraphics[width=.82\columnwidth]{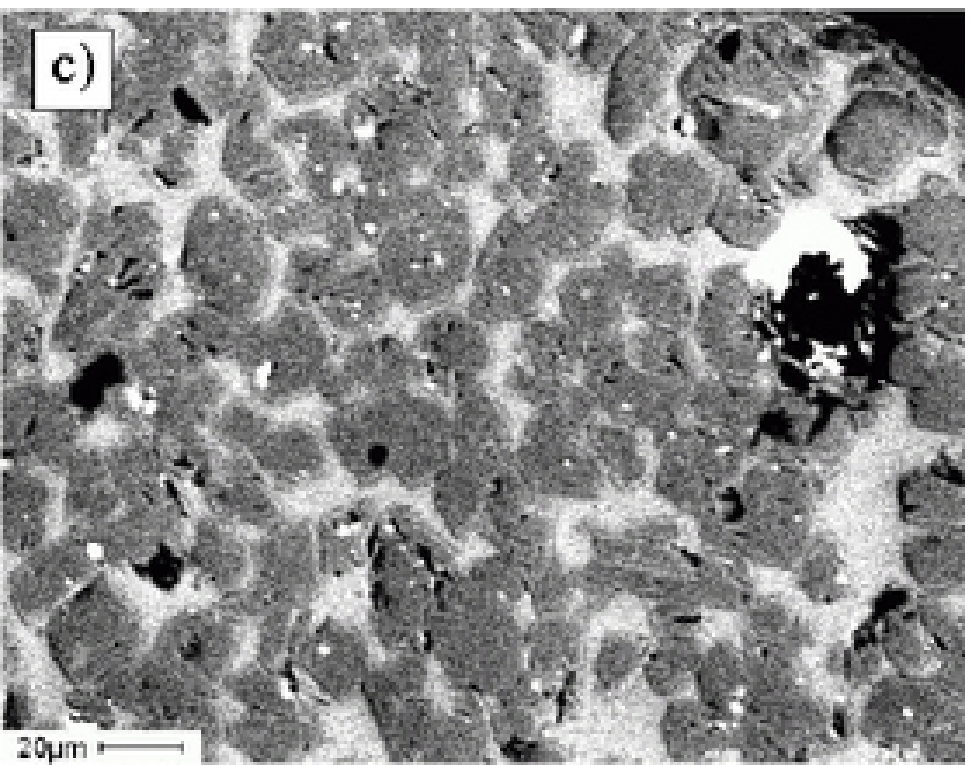}
    \caption{Backscattered electron microscope images of three melted
samples from the synchrotron experiment. (a) cross section of a
complete sample with randomly orientated olivine crystals and type
IA precursor composition. (b) section of a sample with barred
olivine crystals and type II precursor composition. (c) section of
a sample with porphyritic crystals and type IAB precursor
composition. The samples in (a) and (c) have cavities of size $<$
500\,mm. Cavities have been observed \citep{maharaj1994} in other
experimentally produced samples and are caused by trapped gases
and incomplete melting.}
    \label{fig:sync}\vspace*{-0.25cm}
\end{figure}

A total of 24 samples were melted and cooled in the radiation
beam. Backscattered electron microscope images of three samples
are given in Fig.\,\ref{fig:sync}. One sample
(Fig.\,\ref{fig:sync}a) has olivine crystals orientated in a
random pattern whereas the sample in Fig.\,\ref{fig:sync}b has a
barred olivine texture. The sample in figure \ref{fig:sync}c has
porphyritic microstructure.

There was a general tendency for the microstructure of the samples
to reflect the liquidus temperature \citep{yu1998}. Type IA
samples have the highest liquidus temperature of 1692\,\degr C
(Table\,\ref{tab:comp}). The type IA samples were predominantly
porphyritic in texture with smaller crystals while IAB were about
equal mixture of porphyritic and acicular olivine whereas acicular
olivine predominated in type II. The type IA samples with highest
liquidus temperature had more nuclei throughout the partial melt
at the start of cooling on which the crystals could grow
\citep{rubin2000}. The greater number of nucleation sites resulted
in more crystals with smaller dimensions. Type II samples had the
least number of nucleation sites resulting in fewer but larger
crystals. The faster the cooling rate the more imperfect the
crystals that formed. There is not a perfect match between the
microstructure of chondrules (Fig.\,\ref{fig:met}) and the samples
produced in the synchrotron beam (Fig.\,\ref{fig:sync}). The final
texture of samples depends on a range of factors
\citep{rubin2000,connolly1998,lofgren1990} such as the precursor
composition, maximum temperature, rate of cooling and duration of
heating at maximum temperature. Further experiments are needed to
explore a wider range in parameter space to obtain more
observational constraints on the process.

\begin{figure}[t]\centering
\includegraphics[width=\columnwidth]{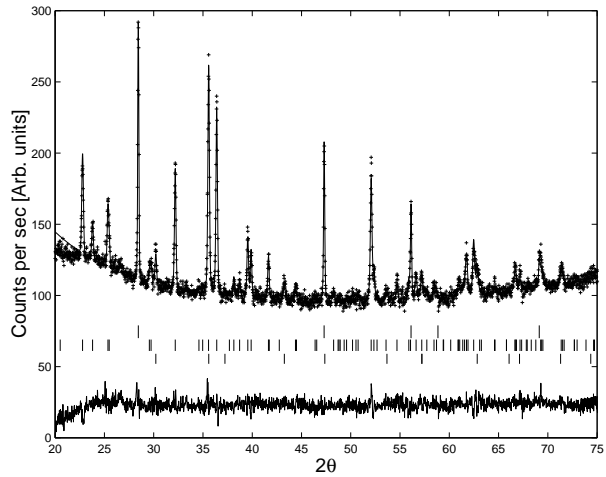}
 \caption{X-ray diffraction pattern of a type II sample. The data
are plotted with crosses and the calculated profile is shown as a
continuous line. The three sets of vertical bars in the middle of
the figure are the calculated reflection positions for silicon
(top), olivine (middle) and magnetite (bottom). The residuals
between the data and the calculated profile are shown at the
bottom of the figure indicating a very good match.}
    \label{fig:reit}
\end{figure}

The Rietveld plot of one type II sample is given in figure
\ref{fig:reit}. The R values, relating the observations to the
model, for type IA samples were typically R$_\mathrm{p}$ = 6.5\,\%
for the profile and R$_\mathrm{wp}$ = 8.2\,\% for the weighted
profile and for type II samples were R$_\mathrm{p}$ = 2.2\,\% and
R$_\mathrm{wp}$ = 2.8\,\% indicating very acceptable fits. The
sample in Fig.\,\ref{fig:sync}a had olivine with weight percentage
of 56.6\,\% with the remainder being amorphous while the type II
sample shown in Fig.\,\ref{fig:reit} had olivine (86.2 wt-\%),
magnetite (7.2 wt-\%) with the remainder being amorphous. The
Bragg R$_\mathrm{B}$ factors for the individual crystalline phases
were also very acceptable with values less than 4\,\%. Refining
the site occupancy factors for the two magnesium and iron sites
allowed the calculation of stoichiometry of the olivine as
Mg$_{1.92}$Fe$_{0.08}$SiO$_{4}$, (i.e. almost pure forsterite) for
the type IA sample (Fig.\,\ref{fig:sync}a) and
Mg$_{1.72}$Fe$_{0.28}$SiO$_{4}$ for the type II sample
(Fig.\,\ref{fig:reit}). The higher iron content of the type II
olivine reflects the greater amount of iron in the starting
mixture.

\section{Discussion}
This experiment demonstrates that GRBs can melt precursor dust
balls to form chondrules in nearby planetary forming systems. Once
formed, the chondrules can move through the gas more freely and
coagulate to form the building blocks of planets
\citep{cuzzi2001}. GRBs with durations greater than 2\,s are
associated with supernovae in massive stars and the formation of
Kerr black holes \citep{popham1999,mcbreen2002}. The discovery of
more than 100 extra-solar giant planets has opened a range of
questions regarding the mechanisms of planetary formation
\citep{santos2003,laughlin2000}. The probability that any
planetary forming system will be blasted by a nearby GRB has been
estimated \citep{mcbreen1999,scalo2002} to be about 0.1\,\%. In
the solar neighbourhood, 7\,\% of stars with high metallicity
harbour a planet whereas less than 1\,\% of stars with solar
metallicity seem to have a planet \citep{santos2003}. The GRB
method seems to be only one of the mechanisms involved because of
the high percentage of stars with planets. There is a further
difficulty for the model in the case of our solar system because
most chondrules were melted more than once requiring a repeating
process \citep{rubin2000,wasson2003}. However the formation of
planetary systems may be enhanced by the presence of a nearby GRB
which will form chondrules across the whole nebula at essentially
the same time \citep{mcbreen1999} but would require a range of
assumptions about the location of the dust in the disk to account
for the properties of chondrules in different classes of
chondrites. Advances in the methods of detecting remnant GRBs and
planetary systems may reveal a link between them in our galaxy.

A GRB can reveal planetary forming systems in other galaxies
because there will be short duration ($\sim$\,1\,hour) bursts of
infrared radiation from the melted dust when chondrules form
across the whole nebula. These infrared bursts can occur for up to
several hundred years after the GRB when the expanding shell of
radiation melts the dust in planetary forming systems. However the
GRB should be in a nearby galaxy to detect the faint infrared
bursts with powerful telescopes such as the Overwhelmingly Large
Telescope and the Next Generation Space Telescope.

\begin{acknowledgements}
The University College Dublin group thank Enterprise Ireland for
support. We thank Alan Rubin for his insightful comments.
\end{acknowledgements}
\small

\bibliographystyle{aa}
\bibliography{Ff191}

\end{document}